\documentclass[lettersize,journal]{IEEEtran}
\usepackage{amsmath,amsfonts, amssymb}
\usepackage{array}
\usepackage[caption=false,font=normalsize,labelfont=sf,textfont=sf]{subfig}
\usepackage{textcomp}
\usepackage{stfloats}
\usepackage{url}
\usepackage{verbatim}
\usepackage{graphicx}
\DeclareGraphicsExtensions{.pdf,.png,.jpg}
\usepackage{cite}
\usepackage{braket}
\usepackage{algorithm}
\usepackage{algpseudocode}
\usepackage{booktabs}
\usepackage{booktabs, multirow, siunitx}

\begin{document}

\title{Techno-Economic Feasibility Analysis of Quantum Key Distribution for Power-System Communications}

\author{Ziqing Zhu,~\IEEEmembership{Member,~IEEE} 
}
\maketitle

\begin{abstract}
The accelerating digitalization and decentralization of modern power systems expose critical communication infrastructures to escalating cyber risks, particularly under emerging quantum computing threats. This paper presents an integrated techno-economic framework to evaluate the feasibility of Quantum Key Distribution (QKD) for secure power-system communications. A stochastic system model is developed to jointly capture time-varying key demand, QKD supply under optical-loss constraints, station-side buffering, and post-quantum cryptography (PQC) fallback mechanisms. Analytical conditions are derived for service-level assurance, including buffer stability, outage probability, and availability bounds. Building on this, two quantitative metrics, including the Levelized Cost of Security (LCoSec) and Cost of Incremental Security (CIS), are formulated to unify capital, operational, and risk-related expenditures within a discounted net-present-value framework. Using IEEE 118-bus, 123-node, and 39-bus test systems, we conduct discrete-event simulations comparing PQC-only, QKD-only, and Hybrid architectures across multiple topologies and service profiles. Results show that Hybrid architectures dominated by QKD significantly reduce key-outage probability and SLA shortfalls, achieving near-unit availability for real-time and confidentiality-critical services. Economic analyses reveal clear breakeven zones where QKD-enhanced deployments become cost-effective, primarily in metropolitan and distribution-level networks under moderate optical loss and buffer sizing. The proposed framework provides a reproducible, risk-aware decision tool for guiding large-scale, economically justified QKD adoption in future resilient power-system infrastructures.
\end{abstract}


\section{Introduction}

Modern power systems are undergoing rapid digitalization, decentralization, and interconnection. As a result, critical operations such as protection, control, and metering increasingly rely on communication networks that must meet stringent requirements for latency, availability, and long-term data confidentiality \cite{Cao2022QKDNetworks}. The emergence of quantum computing further exacerbates these challenges. In particular, the ``Store-Now-Decrypt-Later'' (SNDL) threat model renders sensitive power system data and control messages vulnerable to retrospective attacks, even if they are protected today using classical cryptographic schemes \cite{Cao2022QKDNetworks}. In this context, \emph{Quantum Key Distribution} (QKD) offers a compelling long-term security foundation with three distinct advantages: First, QKD generates high-entropy symmetric keys with information-theoretic security guarantees, independent of computational assumptions \cite{Zhang2022DIQKD}; second, its key generation process is \emph{observable, auditable, and measurable}, enabling integration into SLA management and system monitoring \cite{Cao2022QKDNetworks}; third, with recent advances in photonic chipsets, network interfaces, and standardization, QKD is becoming practically deployable alongside legacy OT/IT infrastructures and complementary to post-quantum cryptography (PQC) \cite{Dolphin2023HybridChipQKD,Yang2024IntercitySPS,Zahidy2024HiDQKD}.

Despite its promise, the techno-economic feasibility of large-scale QKD deployment in power systems remains an open question. Existing studies primarily focus on physical-layer performance, protocol-level analysis, or small-scale pilot networks. Some research addresses trusted-node selection and secure routing. However, these efforts often overlook the structural characteristics of power system operations, including: (i) strict delay and availability constraints for real-time services (e.g., GOOSE, PMU); (ii) bursty and hierarchical traffic patterns that impact key demand; and (iii) the tightly coupled dynamics between QKD supply, key buffer management, and fallback strategies (e.g., PQC-based fallback). There is currently no unified model that captures the interactions among \emph{key demand, QKD supply, station-side buffering, and fallback policies}, nor one that allows direct inference of service reliability metrics or lifecycle security costs. Moreover, existing frameworks do not offer a method to translate \emph{security risks into quantifiable cash flows}, which is essential for cross-architecture economic comparisons. Sensitivity to uncertain parameters (e.g., fiber loss, device cost, quantum threat scenarios) and external conditions (e.g., discount rates, SLA penalties) is also insufficiently explored, limiting the utility of prior work for investment planning or policy design.

To address these gaps, we propose an integrated \emph{techno-economic analysis framework} for QKD deployment in power systems. We begin by formulating a vendor-agnostic stochastic system model that jointly captures (i) time-varying key demand driven by protocol-level refresh and authentication rates across heterogeneous services, and (ii) QKD supply governed by loss-to-rate conversion functions and modulated by station-side buffering. Our model incorporates fallback behavior and offers analytical bounds on buffer underflow and SLA satisfaction conditions. We then introduce two evaluation metrics---\emph{Levelized Cost of Security} (LCoSec) and \emph{Cost of Incremental Security} (CIS)---which integrate CAPEX, OPEX, and quantifiable risk costs into a unified discounted net present value (NPV) framework. By using SLA-adjusted secure output as the normalizing factor, these metrics support meaningful cross-architecture comparisons and threshold-based decision making. We apply the framework to three representative test systems that model metropolitan backbones, distribution-level hubs, and long-haul interconnects. Communication topologies are designed to mirror electrical topologies, and simulations incorporate standard-compliant traffic profiles and disturbance scripts. Under realistic DV and CV QKD performance curves, we conduct unified discrete-event simulations of PQC-only, QKD-only, and hybrid architectures, evaluating availability, delay violations, key exhaustion, and economic metrics. We further conduct comprehensive \emph{sensitivity and scenario robustness analyses}, quantifying the impact of fiber loss, buffer thresholds, refresh rates, pricing structures, and threat scenarios on SLA compliance and lifecycle costs, and identifying the tipping points that shape the cost-benefit landscape.

Our key contributions are threefold: (1) we propose a stochastic system model that links key usage, QKD production, buffering, and fallback strategies, providing quantitative and verifiable conditions for SLA assurance; (2) we construct a \emph{techno-economic evaluation framework} that bridges risk-aware investment cost with SLA-driven secure output, enabling fair benchmarking across security architectures via LCoSec and CIS; and (3) we build a fully reproducible evaluation pipeline, including default parameter sets, topologies, service profiles, scenarios, and analytics, facilitating rapid updating and scenario testing. Experimental results show that hybrid architectures dominated by QKD can absorb supply fluctuations and suppress tail risks in high-real-time and long-duration confidentiality applications. The economic attractiveness of QKD-enhanced deployments is highly sensitive to fiber conditions, buffer design, and key lifecycle policies, with clearly identifiable breakeven zones. These findings offer actionable insights and quantitative tools for rational, risk-aware security planning in the post-quantum era of power systems.

The remainder of this paper is organized as follows. Section 2 reviews related work. Section 3 presents the system model and assumptions. Section 4 formulates the integrated security performance framework. Section 5 introduces the techno-economic model and defines evaluation metrics. Section 6 describes the experimental setup, test systems, and scenario design. Section 7 presents experimental results. Section 8 concludes the paper and outlines directions for future work.

\section{Related Works}

QKD has advanced in performance and scalability in recent years. Twin-field implementations have extended secure transmission distances to the thousand-kilometre regime with ultra low noise detectors \cite{Liu2023PRL,Wang2022NatPhotonics}. Long-term metropolitan operation has been demonstrated on multi node networks, which indicates practical stability on standard optical fiber \cite{Chen2021npjQI134,Paraiso2021NatPhotonics}. Continuous-variable schemes and improved signal processing have raised secret key rates, while photonic integration has delivered chip based transmitters and receivers operating at gigahertz clock rates with real time post processing \cite{Erkilic2025CommunPhys,Chen2021Nature}. Foundational surveys continue to consolidate device level progress and security analyses \cite{Xu2020RMP,Pirandola2020AOP}.

Integration into power system communications has moved from concept to field trials. Utility demonstrations have shown that quantum generated keys can authenticate Supervisory Control and Data Acquisition traffic and support real time grid operations on existing fiber \cite{Alshowkan2022SciRep,Grice2025IEEEAccess}. Pilot testbeds within European projects have connected substations through quantum secured links, which reduces the risk of unauthorized access and control manipulation \cite{Aquina2025EPJQT,Chaturvedi2025SciRep}. These studies suggest that QKD can strengthen confidentiality and integrity for protection, measurement, and control services when combined with appropriate buffering and service level targets \cite{Wang2021npjQI67}.

The techno economic feasibility of wide area deployment depends on throughput, latency, and cost. Recent analyses emphasise evaluation that balances capital and operational expenses against quantifiable risk reduction, defined as avoided losses from cyber incidents \cite{Grice2025IEEEAccess,Chaturvedi2025SciRep}. Network optimisation aims to limit the number of trusted nodes while maintaining coverage across regional grids \cite{Erkilic2025CommunPhys}. In parallel, hybrid cryptographic designs combine QKD for frequent key refresh with post quantum algorithms for authentication and end to end protection. Experiments indicate that such combinations preserve security in the presence of individual component failures and align with defence for critical infrastructure \cite{Garms2024AQT,Cirigliano2024npjQI}.

Growing recognition of the Store Now Decrypt Later threat motivates early adoption of quantum safe communication for high value links. Adversaries may record encrypted operational and market data today and attempt future decryption with quantum computers. QKD addresses this risk since interception on the quantum channel introduces observable errors, which enables timely detection \cite{Xu2020RMP,Pirandola2020AOP}. The literature therefore supports a forward looking approach where QKD is deployed alongside post quantum methods and is guided by risk aware planning for modern power systems \cite{Grice2025IEEEAccess,Alshowkan2022SciRep}.

\section{System Model and Assumptions}
This section presents a unified system model that links power-system applications, cryptographic key demand, QKD key supply and buffering, fallback (PQC) operation, and service-level agreements (SLAs). The model supports both feasibility analysis (availability and delay) and techno-economic comparisons among QKD-only, PQC-only, and hybrid architectures.

\subsection{Network Topology and Traffic Classes}
We consider a three-tier utility communication network with backbone, aggregation, and substation (edge) layers. Let the node set be \(\mathcal{V}\) and the directed link set be \(\mathcal{E}\subseteq\mathcal{V}\times\mathcal{V}\). Each link \(\ell=(i,j)\in\mathcal{E}\) has physical length \(d_\ell\) and total optical loss \(L_\ell\) (dB), which aggregates fiber attenuation and fixed losses (connectors/splices/splitters). Each node \(i\in\mathcal{V}\) aggregates traffic and hosts a local key buffer.

We group power-system traffic into classes by real-time and reliability requirements, e.g., GOOSE/protection, SV (sampled values), PMU/WAMS, SCADA/AGC, and metering/market/settlement/audit. For any class \(k\) at node \(i\), we use the tuple
\begin{align}
(\lambda_{k,i},\, s_k,\, L_k,\, A_k)
\end{align}
where \(\lambda_{k,i}\) (pkt/s) is the arrival rate, \(s_k\) (bits) the payload size per message, \(L_k\) (s) the end-to-end delay bound, and \(A_k\in(0,1]\) the availability target (e.g., \(0.9999\)). To capture long-term confidentiality (SNDL risk), we also specify a confidentiality horizon \(T_k^{\mathrm{conf}}\) (years) for later economic valuation.

For arrival processes, we assume Poisson arrivals for regular telemetry/control (e.g., SCADA/AGC, PMU). Event-driven GOOSE/SV can be bursty; we upper-bound bursts by a Markov-modulated Poisson process (MMPP) surrogate using a burst factor \(\beta_k\ge 1\). Over a discrete time step \(\Delta t\) (we use \(\Delta t=1\) s unless stated), the worst-case instantaneous rate is approximated by
\begin{align}
\hat{\lambda}_{k,i} \;=\; \beta_k\, \lambda_{k,i}.
\end{align}

\subsection{Cryptographic Policy and Key-Demand Model}
We compare: (i) QKD-refreshed symmetric cryptography (e.g., AES-GCM) with message authentication (MAC), (ii) QKD with one-time pad (OTP) for very small, highly sensitive packets, (iii) PQC-only, and (iv) a hybrid where QKD refreshes session keys while PQC provides entity authentication and fallback.

For class \(k\), let \(f_k\) (1/s) denote the session-key refresh frequency, \(\ell_{\mathrm{sess}}\) (bits) the session-key length, and \(\ell_{\mathrm{mac}}\) (bits) the per-message MAC key/label budget. The symmetric-key demand (session refresh + per-message authentication) for class \(k\) at node \(i\) is
\begin{align}
D^{\mathrm{sym}}_{k,i}
&= f_k\, \ell_{\mathrm{sess}} \;+\; \lambda_{k,i}\, \ell_{\mathrm{mac}} \quad (\mathrm{bits/s}),
\end{align}
where the first term accounts for session rekeying and the second for per-message authentication/anti-replay (e.g., Carter--Wegman MACs mapped to a conservative one-time key budget).

If OTP is (sparingly) used for tiny, highly sensitive packets, the key demand equals the plaintext rate:
\begin{align}
D^{\mathrm{OTP}}_{k,i}
&= \lambda_{k,i}\, s_k \quad (\mathrm{bits/s}).
\end{align}
The total key demand at node \(i\) aggregates all traffic classes present at that node:
\begin{align}
D_i
&= \sum_{k\in\mathcal{K}(i)}
\Big(
D^{\mathrm{sym}}_{k,i} \;+\; D^{\mathrm{OTP}}_{k,i}
\Big),
\end{align}
where \(\mathcal{K}(i)\) is the set of traffic classes originated or aggregated at \(i\). In discrete time, the per-step demand is \(d_{i,t}=D_i\,\Delta t\).

\subsection{QKD Supply and Link/Device Model}
Each QKD link \(\ell\in\mathcal{E}\) provides a steady-state key generation rate \(R_\ell\) (bits/s) that depends on total loss \(L_\ell\), distance \(d_\ell\), QBER, and implementation parameters (DV/CV, detector efficiency, clock rate, stabilization). We keep the model protocol-agnostic and use a monotone decreasing parametric form or a fit to measured/vendor curves:
\begin{align}
R_\ell &= g\!\big(L_\ell;\,\theta\big),
&
L_\ell &= \alpha\, d_\ell \;+\; L_{\ell}^{\mathrm{fix}},
\end{align}
where \(\alpha\) (dB/km) is fiber attenuation, \(L_{\ell}^{\mathrm{fix}}\) (dB) captures fixed losses, and \(\theta\) collects system parameters. A common DV-QKD approximation is \(R_\ell \approx R_0\,10^{-\eta L_\ell}\) with \(\eta>0\), but in evaluation we will use measured or published curves. The total QKD supply arriving at node \(j\) is
\begin{align}
A_j
&= \sum_{\ell=(i,j)\in\mathcal{E}} R_\ell,
\end{align}
and the per-step supply is \(a_{j,t}=A_j\,\Delta t\). Keys are landed at the local KMS as “key blocks”; we treat them as a divisible bit pool for tractability (block fragmentation overhead can be engineered away in practice).

\subsection{Key Buffer Dynamics and Fallback Policy}
Each node \(i\) maintains a finite key buffer with capacity \(B_i^{\max}\) (bits) and a minimum safety threshold \(b_i^{\min}\) (bits). In discrete time, buffer evolution follows
\begin{align}
B_{i,t+1}
&= \min\!\Big\{
B_i^{\max},\;
\max\{0,\; B_{i,t} - d_{i,t}\} \;+\; a_{i,t}
\Big\}.
\end{align}
An \emph{out-of-key} (outage) event occurs if \(B_{i,t} < b_i^{\min}\). Upon outage, a preconfigured fallback activates: a proportion \(\phi_i\in[0,1]\) of affected traffic shifts to PQC-only (or a degraded mode with reduced refresh/authentication). Let \(\chi_{i,t}\in\{0,1\}\) indicate fallback at node \(i\) during step \(t\). Defining the global states \(\mathbf{B}_t=(B_{i,t})_{i\in\mathcal{V}}\) and \(\mathbf{X}_t=(\chi_{i,t})_{i\in\mathcal{V}}\), the pair \((\mathbf{B}_t,\mathbf{X}_t)\) forms a Markov process driven by \((a_{i,t},d_{i,t})\). For a given traffic/supply model, the steady-state outage probability at node \(i\) is
\begin{align}
p_i^{\mathrm{out}}
&= \Pr\!\big(B_i < b_i^{\min}\big).
\end{align}
A necessary (first-order) stability condition in expectation is the positive average margin:
\begin{align}
\mathbb{E}[A_i]
&\;\ge\;
\mathbb{E}[D_i] \;+\; \delta_i,
\end{align}
where \(\delta_i>0\) safeguards against stochastic fluctuations and maintenance downtime. If the margin condition is violated, the long-run outage probability tends to one, even if short-term operation appears feasible.

\subsection{Delay and Availability Assumptions}
End-to-end delay is the sum of propagation/forwarding, queuing/shaping, cryptographic processing (including KMS key retrieval and session set-up), and any extra overhead due to fallback switching. For class \(k\),
\begin{align}
\mathrm{Delay}_{k}
&= T^{\mathrm{prop}}
\;+\; T^{\mathrm{queue}}_{k}
\;+\; T^{\mathrm{crypto}}_{k}
\;+\; \chi\, T^{\mathrm{fallback}}_{k},
\end{align}
where \(T^{\mathrm{prop}}\) is determined by topology and distance, \(T^{\mathrm{queue}}_{k}\) by traffic shaping/queuing, \(T^{\mathrm{crypto}}_{k}\) by crypto/KMS operations, \(\chi\in\{0,1\}\) the fallback indicator, and \(T^{\mathrm{fallback}}_{k}\) the additional negotiation/switching overhead. In normal (non-fallback) operation, \(T^{\mathrm{prop}}\) and \(T^{\mathrm{crypto}}_{k}\) are assumed bounded with small variance; meeting the SLA \(\Pr(\mathrm{Delay}_k \le L_k)\ge A_k\) therefore hinges on controlling the outage probability and the share of time in fallback.

We approximate class-\(k\) availability as the fraction of time the service meets its delay bound and cryptographic function is operational:
\begin{align}
\mathrm{Availability}_k
&\approx
1
\;-\; \Pr\!\big(B < b^{\min}\big)
\;-\; \Pr\!\big(\mathrm{Delay}_k > L_k \,\big|\, \chi=1\big)\,
\Pr(\chi=1).
\end{align}
The first subtraction accounts for out-of-key outages; the second captures delay violations specifically in fallback, weighted by the fallback occupancy.

\subsection{Parameterization and Modeling Scope}
To ensure reproducibility, parameters are split into default values and sensitivity ranges. Network parameters \((\alpha, L^{\mathrm{fix}}, d_\ell)\) come from fiber plant records; traffic parameters \((\lambda_{k,i}, s_k, L_k, A_k, \beta_k, f_k, \ell_{\mathrm{sess}}, \ell_{\mathrm{mac}})\) come from utility operating rules or measurements; QKD parameters are provided via the fitted function \(g(\cdot;\theta)\) from lab/field data. Scope limitations include: (i) treating keys as a divisible bit pool (block fragmentation overhead is neglected but can be engineered to be negligible), (ii) absorbing DV/CV and impairment statistics into \(g(\cdot)\) rather than modeling protocol specifics, and (iii) not explicitly modeling clock/synchronization and temperature-induced drifts (covered indirectly via the safety margin \(\delta_i\) and sensitivity scenarios).

The above model underpins four subsequent steps: (1) key supply–demand and outage probability estimation at node/link level; (2) delay/availability evaluation against SLAs; (3) embedding total cost of ownership and risk into a levelized security metric; and (4) deployment optimization over trust-node placement, buffer sizing, and fallback tuning.

\section{Stochastic Security Model for QKD-Enabled Grid Communications}
This section develops a single stochastic framework that ties together (i) cryptographic key demand induced by power-system applications, (ii) QKD-driven key supply with finite buffering and PQC fallback, and (iii) service-level outcomes such as delay and availability. Time is discretized with step \(\Delta t\) (default \(\Delta t=1\,\mathrm{s}\)); per-second quantities map to per-step values by multiplication with \(\Delta t\).

\subsection{Keying Process and Demand}
Consider node \(i\) and traffic class \(k\). Let \(\lambda_{k,i}\) (pkt/s) denote the message arrival rate, \(s_k\) (bits) the payload per message, \(f_k\) (1/s) the session-key refresh frequency, \(\ell_{\mathrm{sess}}\) (bits) the session-key length, and \(\ell_{\mathrm{mac}}\) (bits) the per-message authentication budget (e.g., Carter--Wegman style). Under symmetric cryptography (e.g., AES-GCM with MAC), the per-second key demand decomposes into session refresh and per-message authentication:
\begin{align}
D^{\mathrm{sym}}_{k,i}
&= f_k\,\ell_{\mathrm{sess}}
\;+\;
\lambda_{k,i}\,\ell_{\mathrm{mac}}
\quad (\mathrm{bits/s}) .
\end{align}
If one-time pad (OTP) is used for rare, ultra-sensitive tiny packets, the key demand equals the plaintext rate:
\begin{align}
D^{\mathrm{OTP}}_{k,i}
&= \lambda_{k,i}\, s_k
\quad (\mathrm{bits/s}) .
\end{align}
Aggregating over all classes \(\mathcal{K}(i)\) present at node \(i\) yields
\begin{align}
D_i
&= \sum_{k\in\mathcal{K}(i)}
\Big(
D^{\mathrm{sym}}_{k,i}
\;+\;
D^{\mathrm{OTP}}_{k,i}
\Big),
&
d_{i,t}
&= D_i\,\Delta t .
\end{align}
To conservatively account for burstiness (e.g., GOOSE/SV), we introduce a burst factor \(\beta_k\ge 1\) and bound the instantaneous rate by \(\hat{\lambda}_{k,i}=\beta_k\lambda_{k,i}\). A robust demand envelope is then
\begin{align}
\hat{D}_i
&=
\sum_{k\in\mathcal{K}(i)}
\Big(
f_k\,\ell_{\mathrm{sess}}
\;+\;
\hat{\lambda}_{k,i}\,\ell_{\mathrm{mac}}
\;+\;
\lambda_{k,i}\, s_k^{(\mathrm{OTP})}
\Big),
\end{align}
where \(s_k^{(\mathrm{OTP})}=s_k\) only for classes that actually employ OTP and is \(0\) otherwise.

\subsection{QKD-Driven Supply and Buffering}
For each fiber link \(\ell=(j\!\to\! i)\), the total optical loss combines propagation and fixed penalties:
\begin{align}
L_\ell
&= \alpha\, d_\ell \;+\; L_{\ell}^{\mathrm{fix}}
\quad (\mathrm{dB}) ,
\end{align}
with \(\alpha\) (dB/km) the attenuation coefficient, \(d_\ell\) (km) the physical length, and \(L_{\ell}^{\mathrm{fix}}\) (dB) connector/splice/splitter losses. The steady-state key generation rate is modeled in a protocol-agnostic fashion via a decreasing fit to lab/field data,
\begin{align}
R_\ell
&= g\!\big(L_\ell;\,\theta\big)
\quad (\mathrm{bits/s}) ,
\end{align}
where \(\theta\) collects implementation parameters (DV/CV, detectors, clock, stabilization). The aggregate per-second supply arriving at node \(i\) is
\begin{align}
A_i
&= \sum_{\ell=(j,i)} R_\ell,
&
a_{i,t}
&= A_i\,\Delta t .
\end{align}
When links feed multiple buffers, we may introduce per-step allocation variables \(u_{\ell\to i,t}\in[0,\,R_\ell\Delta t]\) with \(\sum_i u_{\ell\to i,t}\le R_\ell\Delta t\) and set \(a_{i,t}=\sum_{\ell} u_{\ell\to i,t}\). This captures max--min or SLA-weighted sharing across nodes.

Each node \(i\) maintains a finite key buffer with capacity \(B_i^{\max}\) (bits) and a minimum safety threshold \(b_i^{\min}\) (bits). The buffer evolves as a reflected, bounded random walk:
\begin{align}
B_{i,t+1}
&=
\min\!\Big\{
B_i^{\max},\;
\max\{0,\, B_{i,t} - d_{i,t}^{\mathrm{eff}}\}
\;+\; a_{i,t}
\Big\}.
\end{align}
If \(B_{i,t}<b_i^{\min}\), an out-of-key event triggers a fallback where a fraction \(\phi_i\in[0,1]\) of affected traffic transitions to PQC-only (or a degraded mode), thereby instantaneously reducing key consumption. With a fallback indicator \(\chi_{i,t}=\mathbb{I}\{B_{i,t}<b_i^{\min}\}\), the effective per-step demand becomes
\begin{align}
d_{i,t}^{\mathrm{eff}}
&=
\big(1-\phi_i\,\chi_{i,t}\big)\, d_{i,t}.
\end{align}
Define the net increment \(X_{i,t}=a_{i,t}-d_{i,t}^{\mathrm{eff}}\). Under standard stationarity assumptions on \(\{X_{i,t}\}\), the coupled process \(\{(B_{i,t},\chi_{i,t})\}\) is Markovian. A first-order stability requirement in expectation enforces a positive margin:
\begin{align}
\mathbb{E}[A_i]
&\;\ge\;
\mathbb{E}[D_i]
\;+\;
\delta_i ,
\end{align}
with \(\delta_i>0\) absorbing key-rate fluctuations, calibrations/maintenance downtime, and traffic bursts. If the margin condition fails persistently, the steady-state out-of-key probability approaches one.

When \(\{X_{i,t}\}\) is i.i.d. and admits a cumulant generating function \(\Lambda_{X_i}(\theta)=\log \mathbb{E}[e^{\theta X_{i,t}}]\), a Cram\'er--Lundberg large-deviation bound yields an exponential tail for underflow. Let \(\kappa_i>0\) solve \(\Lambda_{X_i}(-\kappa_i)=0\). Then, for regimes where capacity does not dominate,
\begin{align}
\Pr\!\big(B_{i,t}<b_i^{\min}\big)
&\;\lesssim\;
C_i\, \exp\!\big(-\kappa_i\, b_i^{\min}\big),
\end{align}
with constant \(C_i\) depending on initialization and reflection. The bound highlights the exponential suppression of outage by increasing \(b_i^{\min}\), enlarging the mean margin, or reducing the variance of \(X_{i,t}\). In practice we use this as a fast screen and calibrate it via Monte Carlo or Markov-chain solvers.

\subsection{Service-Level Metrics and Reliability}
End-to-end delay for class \(k\) is modeled as the sum of propagation/forwarding, queuing/shaping, cryptographic processing (including KMS key retrieval and session setup), and any additional overhead due to fallback switching:
\begin{align}
\mathrm{Delay}_{k}
&=
T^{\mathrm{prop}}
\;+\;
T^{\mathrm{queue}}_{k}
\;+\;
T^{\mathrm{crypto}}_{k}
\;+\;
\chi\, T^{\mathrm{fallback}}_{k}.
\end{align}
In non-fallback operation, \(T^{\mathrm{prop}}\) and \(T^{\mathrm{crypto}}_{k}\) are assumed bounded with small variance; achieving the SLA \(\Pr(\mathrm{Delay}_k \le L_k)\ge A_k\) therefore hinges on suppressing out-of-key events and limiting the fraction of time spent in fallback.

We approximate the availability for class \(k\) as the fraction of time during which the service is both cryptographically operational and within its delay bound:
\begin{align}
\mathrm{Availability}_k
&\approx
1 - \Pr\!\big(B<b^{\min}\big) \nonumber\\
&\quad
- \Pr\!\big(\mathrm{Delay}_k>L_k \,\big|\, \chi=1\big)\,
  \Pr(\chi=1).
\end{align}
The first subtraction term captures outright cryptographic outages; the second accounts for delay violations specifically while in fallback, weighted by the fallback occupancy. A conservative bound can separate queuing and switching contributions. For any threshold \(\tau_k>0\),
\begin{align}
p_{k,\mathrm{fb}}
&\le
\Pr\!\big(T^{\mathrm{queue}}_{k}>\tau_k\big)
+\Pr\!\Big(T^{\mathrm{fb}}_{k}>L_k-\tau_k-T^{\mathrm{prop}}-T^{\mathrm{crypto}}_{k}\Big),
\qquad \tau_k>0,
\end{align}
The first term may be estimated from the network shaping/queueing regime under load, while the second follows from implementation or vendor bounds for negotiation and parameter switching. For hard real-time classes (e.g., protection/GOOSE), engineering practice often chooses \(\phi_i\) and redundant local policies such that fallback keeps \(T^{\mathrm{fallback}}_{k}\) within a tight, certified envelope.

Together, the expectation-margin condition \(\mathbb{E}[A_i]\ge \mathbb{E}[D_i]+\delta_i\), the exponential underflow bound \(\Pr(B<b^{\min})\lesssim C_i e^{-\kappa_i b^{\min}}\), and the availability decomposition above provide three complementary, checkable levers for design: the first screens feasible supply curves, placements, and refresh strategies; the second guides the safety threshold and buffer sizing; the third isolates supply-side outages from fallback-induced delay risk, enabling targeted improvements on both sides to meet per-class SLAs.

\section{Techno\textendash Economic Model}
Building on the stochastic security model, this section specifies a technology\textendash
economic framework that fairly compares \emph{PQC-only}, \emph{QKD-only}, and
\emph{Hybrid} architectures. We cash-flow all expenditures over a finite horizon,
discount them to present value, and normalize by a discounted measure of
\emph{effective security output}. The resulting metric, the \emph{Levelized Cost of
Security} (LCoSec), supports apples-to-apples evaluation across architectures and
traffic mixes.

\subsection{Cost Composition and Cash Flows}
Let the evaluation horizon be discrete years \(t=0,1,\dots,T\), and let \(r>0\) be
the discount rate. Positive cash flows represent expenditures (costs), while
negative terms capture recoveries (e.g., salvage). Denote annual capital
expenditures, operating expenditures, and risk costs by
\(\mathrm{CAPEX}_t,\ \mathrm{OPEX}_t,\ \mathrm{Risk}_t\), respectively. The present value of
total costs is
\begin{align}
\mathrm{NPV}_{\mathrm{cost}}
&=
\sum_{t=0}^{T}
\frac{\mathrm{CAPEX}_t + \mathrm{OPEX}_t + \mathrm{Risk}_t}{(1+r)^t}
\;-\;
\frac{S_T}{(1+r)^T},
\end{align}
where \(S_T\le 0\) is the terminal salvage (a negative cash flow indicating
recovery). For annualized comparison, the equivalent annual cost (EAC) is
\begin{align}
\mathrm{EAC}
&=
\mathrm{NPV}_{\mathrm{cost}}
\cdot
\frac{r(1+r)^T}{(1+r)^T-1}.
\end{align}

Operating expenditures are the sum of energy, lease, maintenance/calibration,
staffing/training, compliance upkeep, and spares. For illustration, if
\(\mathcal{H}_t\) is the set of devices in service during year \(t\), with average
power \(P_h\) (kW), tariff \(e_t\) (currency/kWh), and effective operating hours
\(H_t\), then
\begin{align}
\mathrm{OPEX}^{(\mathrm{energy})}_t
&=
\sum_{h\in\mathcal{H}_t}
P_h \, H_t \, e_t .
\end{align}
For fiber leases, with link length \(d_\ell\) and annual unit price
\(c^{(\mathrm{lease})}_{\ell,t}\),
\begin{align}
\mathrm{OPEX}^{(\mathrm{lease})}_t
&=
\sum_{\ell}
c^{(\mathrm{lease})}_{\ell,t}\, d_\ell .
\end{align}
An aggregated maintenance proxy may be modeled as a fraction \(\mu_t\) of the
in-service capital base:
\begin{align}
\mathrm{OPEX}^{(\mathrm{O\&M})}_t
&=
\mu_t \cdot
\Bigg(
\sum_{\tau=0}^{t}
\mathrm{CAPEX}_\tau \, \rho_{\tau\to t}
\Bigg),
\end{align}
where \(\rho_{\tau\to t}\in[0,1]\) is the fraction of assets from year \(\tau\) that
remain in service in year \(t\) (lifetime/retirement profile).

Risk costs map reliability and confidentiality hazards to expected monetary
losses. First, \emph{SLA shortfall and downtime} costs aggregate per-class
penalties proportional to unavailability. With per-class weights \(Q_{k,t}\)
(e.g., SLA-weighted message volume, protected MWh, or transactions) and
monetary penalty coefficients \(c^{(\mathrm{sla})}_{k,t}\),
\begin{align}
\mathrm{Risk}^{(\mathrm{SLA})}_t
&=
\sum_{k}
c^{(\mathrm{sla})}_{k,t}\,
\Big[1-\mathrm{Availability}_{k,t}\Big]\,
Q_{k,t}.
\end{align}
Second, \emph{store-now-decrypt-later (SNDL)} costs quantify the expected loss
from long-horizon confidentiality breaches. Let \(V^{(\mathrm{conf})}_{k,t}\) be the
annualized value-at-risk for class \(k\) in year \(t\), and let
\(p^{(\mathrm{br})}_{k,t}(a)\) be the hazard (conditional breach probability) under
architecture \(a\in\{\mathrm{PQC},\mathrm{QKD},\mathrm{Hybrid}\}\). Then
\begin{align}
\mathrm{Risk}^{(\mathrm{SNDL})}_t
&=
\sum_{k}
p^{(\mathrm{br})}_{k,t}(a)\;
V^{(\mathrm{conf})}_{k,t},
\\
\mathrm{Risk}_t
&=
\mathrm{Risk}^{(\mathrm{SLA})}_t
+
\mathrm{Risk}^{(\mathrm{SNDL})}_t .
\end{align}
In practice, \(p^{(\mathrm{br})}_{k,t}(\mathrm{QKD})\) is modeled as a small residual
implementation/operations risk, whereas
\(p^{(\mathrm{br})}_{k,t}(\mathrm{PQC})\) evolves with cryptanalytic progress and the
quantum threat timeline.

\subsection{Security Output and the LCoSec Metric}
To normalize across architectures and traffic mixes, we define the
\emph{effective security output} as the discounted sum of SLA-qualified service
value. Let \(q_{k,t}\) be the raw service measure for class \(k\) in year \(t\) (e.g.,
messages or bits within scope, protected MWh, or cleared transactions), and
\(w_k>0\) be an importance weight. The SLA-adjusted annual output is
\begin{align}
\mathrm{SecVal}_t
&=
\sum_{k}
w_k\,
\mathrm{Availability}_{k,t}\,
q_{k,t},
\end{align}
and its present value is
\begin{align}
\mathrm{PV}(\mathrm{SecVal})
&=
\sum_{t=0}^{T}
\frac{\mathrm{SecVal}_t}{(1+r)^t}.
\end{align}
The \emph{Levelized Cost of Security} for architecture \(a\) is
\begin{align}
\mathrm{LCoSec}(a)
&=
\frac{
\mathrm{NPV}_{\mathrm{cost}}(a)
}{
\mathrm{PV}(\mathrm{SecVal})(a)
}
\quad
\end{align}
Class-resolved views may also be reported via
\(\mathrm{LCoSec}_k=\mathrm{NPV}_{\mathrm{cost}}/
\sum_t \mathrm{SecVal}_{k,t}/(1+r)^t\).

\subsection{Architectural Comparisons and Break\textendash Even Conditions}
Fix a baseline architecture \(a_0\) (typically PQC-only). For any candidate
architecture \(a\),
\begin{align}
\Delta \mathrm{NPV}(a)
&=
\mathrm{NPV}_{\mathrm{cost}}(a)
-
\mathrm{NPV}_{\mathrm{cost}}(a_0),
\\
\Delta \mathrm{PV}(\mathrm{SecVal})(a)
&=
\mathrm{PV}(\mathrm{SecVal})(a)
-
\mathrm{PV}(\mathrm{SecVal})(a_0).
\end{align}
If \(\Delta \mathrm{PV}(\mathrm{SecVal})(a)>0\), the \emph{Cost of Incremental Security}
(CIS) is
\begin{align}
\mathrm{CIS}(a\mid a_0)
&=
\frac{
\Delta \mathrm{NPV}(a)
}{
\Delta \mathrm{PV}(\mathrm{SecVal})(a)
}.
\end{align}
Alternatively, introduce an internal shadow price \(\pi>0\) for a unit of
effective security (derived from historical losses, regulatory penalties, or
resilience valuation). Architecture \(a\) has nonnegative net benefit over
\(a_0\) when
\begin{align}
\pi \cdot
\Delta \mathrm{PV}(\mathrm{SecVal})(a)
&\;\ge\;
\Delta \mathrm{NPV}(a).
\end{align}

\subsection{Linking Reliability to Cost and Hybrid Occupancy}
From the stochastic model, availability admits the decomposition
\(\mathrm{Availability}_k
\approx
1 - \Pr(B<b^{\min})
- \Pr(\mathrm{Delay}_k > L_k \mid \chi=1)\Pr(\chi=1)\).
Thus, buffer thresholds \(b^{\min}\), supply margin \(\delta\), and fallback
occupancy \(\Pr(\chi=1)\) enter directly into
\(\mathrm{Risk}^{(\mathrm{SLA})}_t\) via \(\mathrm{Availability}_{k,t}\), quantifying the
economic value of improving key rate, buffer sizing, or fallback procedures.

For \emph{Hybrid} operation, let \(\rho_{k,t}\in[0,1]\) denote the fraction of
time (or traffic) for which class \(k\) benefits from QKD-supported keys in year
\(t\). A convex-hazard approximation yields
\begin{align}
p^{(\mathrm{br})}_{k,t}(\mathrm{Hybrid})
&\approx
(1-\rho_{k,t})\,
p^{(\mathrm{br})}_{k,t}(\mathrm{PQC})
\;+\;
\rho_{k,t}\,
p^{(\mathrm{br})}_{k,t}(\mathrm{QKD}),
\end{align}
with \(p^{(\mathrm{br})}_{k,t}(\mathrm{QKD})\) a small residual reflecting
implementation/operations risks. Because \(\rho_{k,t}\) is determined by the
supply\textendash buffer\textendash fallback dynamics, the SNDL and SLA risk channels close
consistently into \(\mathrm{Risk}_t\).

\subsection{Scenario Uncertainty and Robust Evaluation}
Prices, energy tariffs, lease rates, discount rates, key-rate curves, and threat
levels are uncertain. Let \(\Omega\) be a scenario set with probabilities
\(\{\pi_\omega\}\), and denote by
\(\mathrm{NPV}_{\mathrm{cost}}^{(\omega)}(a)\) and
\(\mathrm{PV}(\mathrm{SecVal})^{(\omega)}(a)\) the corresponding present values in
scenario \(\omega\). An expected LCoSec and chance-constrained SLA specification
are
\begin{align}
\mathbb{E}\big[\mathrm{LCoSec}(a)\big]
&=
\sum_{\omega\in\Omega}
\pi_\omega\,
\frac{
\mathrm{NPV}_{\mathrm{cost}}^{(\omega)}(a)
}{
\mathrm{PV}(\mathrm{SecVal})^{(\omega)}(a)
},
\\
\Pr_\omega\!\Big(
\mathrm{Availability}_{k,t}^{(\omega)} \ge A_k,\ \forall k,t
\Big)
&\;\ge\; 1-\epsilon .
\end{align}
In practice, one may also impose value-at-risk or CVaR summaries on
\(\mathrm{NPV}_{\mathrm{cost}}\) and \(\mathrm{PV}(\mathrm{SecVal})\) to obtain conservative
upper bounds for \(\mathrm{LCoSec}\) and investment thresholds.

\medskip
The above techno\textendash economic model maps the stochastic
supply\textendash buffer\textendash fallback outcomes into cash flows and provides normalized,
comparable metrics (LCoSec and CIS). In the case studies, we will instantiate
these formulas with realistic parameter tables, generate Monte Carlo estimates of
\(\mathrm{Availability}_{k,t}\), and report Pareto fronts and sensitivity analyses to
derive actionable deployment guidance.

\section{Experimental Design}
This section specifies the experimental setup used to evaluate three security architectures—\emph{PQC-only}, \emph{QKD-only}, and \emph{Hybrid}—under identical operating conditions. We describe the test power systems and their induced communication topologies, traffic generation and service-level targets, cryptographic policies per architecture, QKD supply parameterization with finite key buffers, disturbance scripting, simulation regimen, and the technical/economic metrics to be reported. 

We adopt three representative grid--communications scenarios to span urban backbone, distribution aggregation, and long-haul inter-area use cases. On the power side, we use the IEEE 118-bus system to emulate a ``control center $\rightarrow$ substation'' metropolitan backbone, the IEEE 123-node feeder to represent a hierarchical distribution setting with $4$--$8$ neighborhood aggregation points terminating at a station, and the IEEE 39-bus system to represent two control areas connected by a long-distance intertie. Each electrical node is mapped to a communications site by a ``co-located in substation; station-to-station shortest optical path'' rule so that the resulting communications topology is geometrically consistent while remaining agnostic to electrical power flows. For the metropolitan case, we construct an OPGW/leased-fiber ring with chordal links (a near-mesh); for the distribution case, we obtain a tiered star/tree from neighborhoods to an aggregation tier and onward to the station; for the long-haul case, we place $0$--$3$ trusted (or regeneration) nodes along a $150$--$300$~km equivalent path, yielding a chain or dual-chain structure. Every link $\ell$ is assigned a physical length $d_\ell$ and fixed loss component, while fiber attenuation $\alpha$ and connector/splitter penalties follow default values with sensitivity ranges to ensure comparability across scenarios.

Traffic demand covers operational/business classes M1--M5 (metering upload, market/bid clearing interaction, dispatch commands and acknowledgments, settlement/reconciliation, and audit/archival) and protection/measurement classes (GOOSE/SV and PMU/WAMS). SCADA and PMU streams are generated as Poisson processes with per-node class rates $\lambda_{k,i}$; GOOSE/SV exhibit burstiness modeled via a Markov-modulated Poisson process with a tunable burst factor to control the peak-to-mean ratio. For each class, message size $s_k$, delay bound $L_k$, availability target $A_k$, and confidentiality horizon $T_k^{\mathrm{conf}}$ are drawn from standards-informed intervals and field practice. Cryptographic policies are fixed by architecture: \emph{PQC-only} uses a post-quantum key agreement plus symmetric cipher baseline; \emph{QKD-only} employs ``QKD-refreshed session keys + symmetric encryption + MAC,'' with one-time pad (OTP) permitted only for rare, tiny payloads such as audit or key-management triggers; \emph{Hybrid} uses QKD for session-key refresh and PQC for entity authentication and fallback. Session refresh frequency $f_k$, session-key length $\ell_{\mathrm{sess}}$, and per-message MAC budget $\ell_{\mathrm{mac}}$ are set to common defaults and included in sensitivity sweeps; OTP is restricted to ultra-low-volume flows to provide an upper-bound reference.

On the supply side, each fiber link’s total optical loss is $L_\ell=\alpha d_\ell + L_\ell^{\mathrm{fix}}$, and the per-link key generation rate is provided by a loss--rate curve family $R_\ell=g(L_\ell;\theta)$. We include one representative DV-QKD curve and one representative CV-QKD curve (both fitted to laboratory or public field data) to contrast physical implementations while holding the modeling interface invariant. Each station hosts a finite-capacity key buffer with a maximum $B_i^{\max}$ and a minimum safety threshold $b_i^{\min}$; a supply margin parameter absorbs key-rate variability and maintenance windows. The key-management system (KMS) is approximated as converting block arrivals into a divisible bit pool for the crypto stack. When a link feeds multiple nodes, per-step key allocation across nodes is performed using either max--min fairness or a static SLA-weighted proportion to study how resource contention influences availability of the most stringent classes; allocation rules are fixed \emph{a priori} and are not optimized.

Comparative experiments include exactly three baselines: \emph{PQC-only}, \emph{QKD-only}, and \emph{Hybrid}. All runs share the same traffic scripts, physical-topology scripts, price scripts, and disturbance scripts, and all architectures use identical default parameters where applicable to ensure fairness. Simulations are discrete-event with a $1$~s step; each run spans $30$--$90$ days of representative seasonal load, with the first $24$~hours treated as warm-up and excluded from statistics. To obtain annual-scale economic figures, SLA and occupancy statistics from the simulated window are extrapolated to a year and combined with month-varying tariffs, lease rates, and maintenance windows. Each scenario is replicated with multiple random seeds (Monte Carlo) and reported with $95\%$ confidence intervals.

Disturbances are injected through three script types: \emph{key-rate outages} (production windows down for calibration or planned maintenance), \emph{loss steps} (step increases in $L_\ell$ modeling fiber aging or localized temperature effects), and \emph{link cuts} (temporary fiber interruption). Frequencies and durations follow empirical distributions at the monthly/annual scale. To avoid conclusions hinging on a single geometry or loss regime, each of the three communications topologies is exercised under three attenuation bands—low-loss urban, medium, and high-loss long-haul—and rerun end-to-end.

Evaluation proceeds on a common sampling basis for technical and economic metrics. The technical side reports node-level out-of-key probability, per-class availability, fallback occupancy, and end-to-end delay distributions with exceedance probabilities. The economic side reports equivalent annual cost (EAC), a decomposition of risk costs into SLA- and SNDL-related components, the discounted effective security output, and the resulting LCoSec and CIS values. Sensitivity analyses vary, singly or in pairs, fiber attenuation and fixed losses, session refresh and MAC budgets, buffer capacity and minimum threshold, supply margin and burst factor, QKD curve parameters, and price variables (energy, lease, and discount rate). Robustness is assessed over a scenario set by computing chance-constraint satisfaction for per-class SLA targets and expected/quantile summaries for LCoSec. To support reproducibility, we will release default parameter tables, topology and traffic scripts, and the post-processing code used to generate figures; experiment logs record seeds, versions, and timestamps to ensure consistent cross-platform replication.

\section{Results and Discussions}
\subsection{Reliability and SLA Performance}
\begin{figure}[t]
  \centering
  \includegraphics[width=\linewidth]{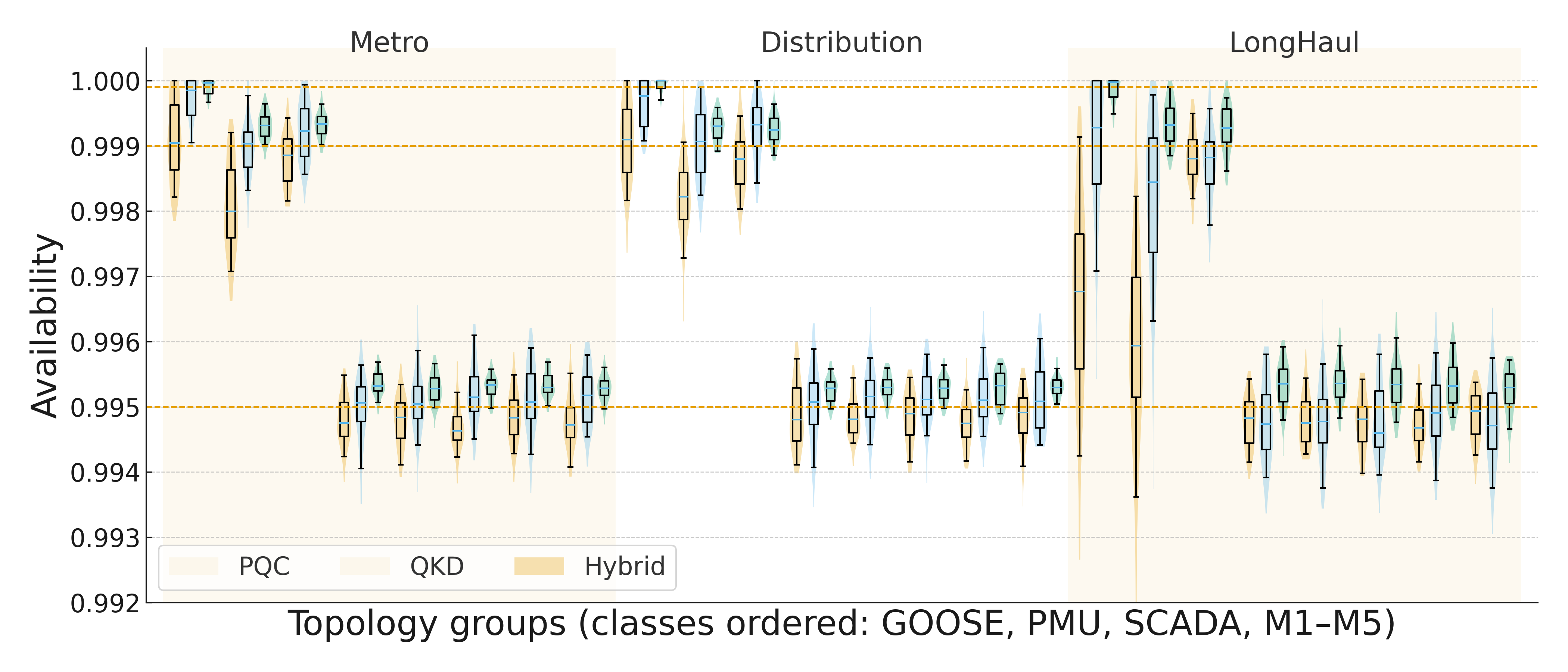}
  \caption{Availability distributions with violin+box overlays. The x-axis is grouped by topology (Metro, Distribution, LongHaul) using shaded bands; within each group, classes are ordered as GOOSE, PMU, SCADA, M1--M5. Dashed horizontal lines indicate representative SLA targets.}
  \label{fig:avail_violin}
\end{figure}

The distributional evidence in Fig.~\ref{fig:avail_violin} indicates that the \emph{Hybrid} architecture consistently dominates both \emph{PQC-only} and \emph{QKD-only} in availability across classes and topologies. The gains are largest for hard real-time traffic over long-haul links. For LongHaul\,$\times$\,GOOSE, the median availability improves from approximately \(0.9969\) under \emph{PQC} to \(0.9990\) under \emph{QKD}, and to essentially unit availability under \emph{Hybrid}; for LongHaul\,$\times$\,PMU the medians move from \(\approx 0.9960\) (\emph{PQC}) to \(\approx 0.9977\) (\emph{QKD}) and \(\approx 0.9993\) (\emph{Hybrid}). In Metro and Distribution settings, all three architectures concentrate tightly around their SLA thresholds, but \emph{Hybrid} compresses the lower tail, signaling fewer extreme shortfalls.

\begin{figure}[t]
  \centering
  \includegraphics[width=\linewidth]{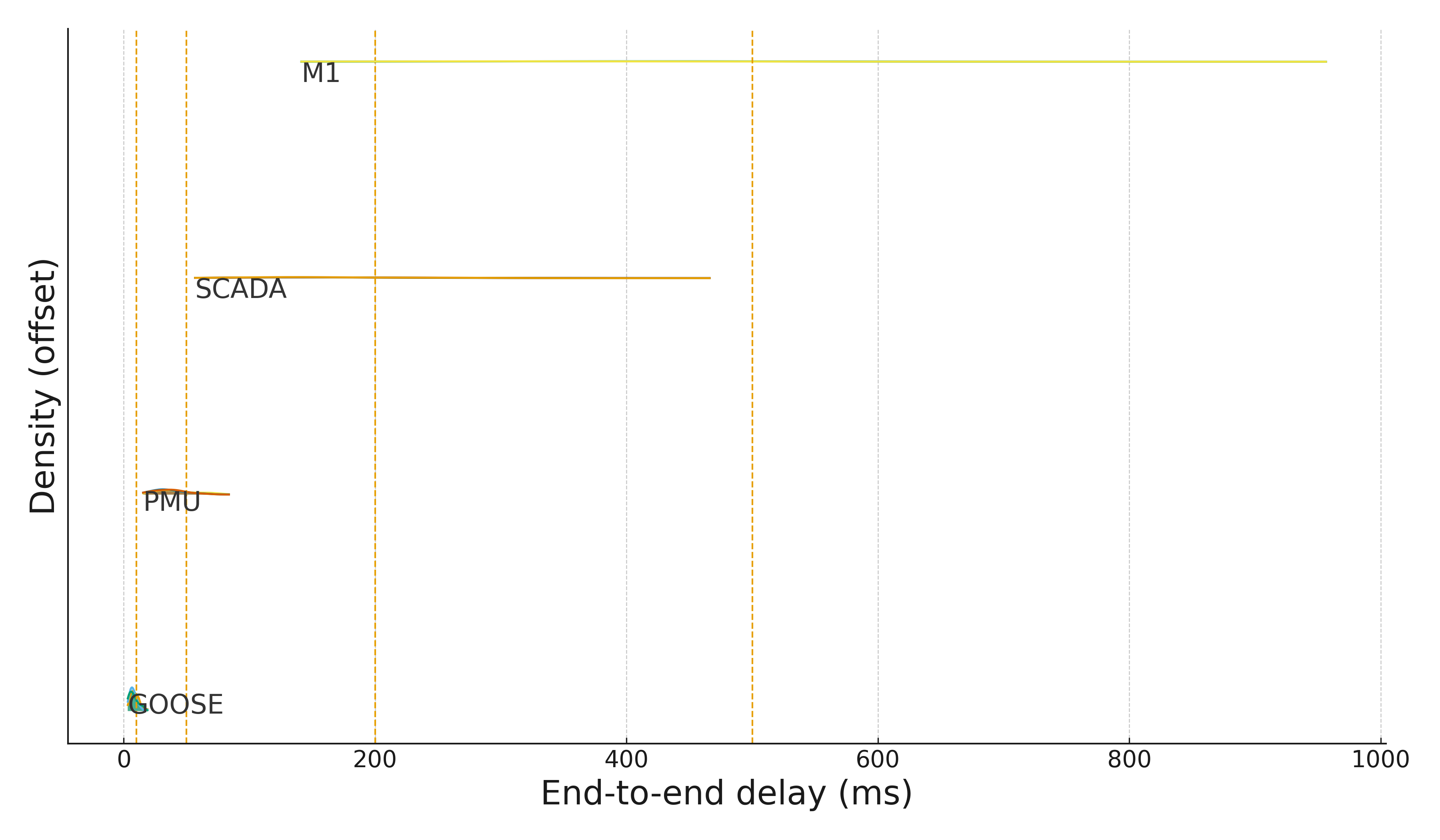}
  \caption{Ridgeline-style end-to-end delay densities for the Metro topology across representative classes (GOOSE, PMU, SCADA, M1); vertical dashed lines mark class-specific delay bounds.}
  \label{fig:delay_ridge}
\end{figure}

Figure~\ref{fig:delay_ridge} corroborates these findings by comparing end-to-end delay shapes on the Metro topology. Exceedance probabilities \( \Pr(\mathrm{Delay}>L_k)\) are reduced under \emph{Hybrid} relative to \emph{PQC} for all shown classes (GOOSE: \(0.15\) vs.\ \(0.22\); PMU: \(0.13\) vs.\ \(0.32\); SCADA: \(0.16\) vs.\ \(0.30\); M1: \(0.15\) vs.\ \(0.33\)). The \emph{QKD-only} curves typically center below \emph{PQC}—consistent with QKD-refreshed symmetric sessions—but present broader shoulders, reflecting occasional key-supply variability when no fallback is present.

\begin{figure}[t]
  \centering
  \includegraphics[width=\linewidth]{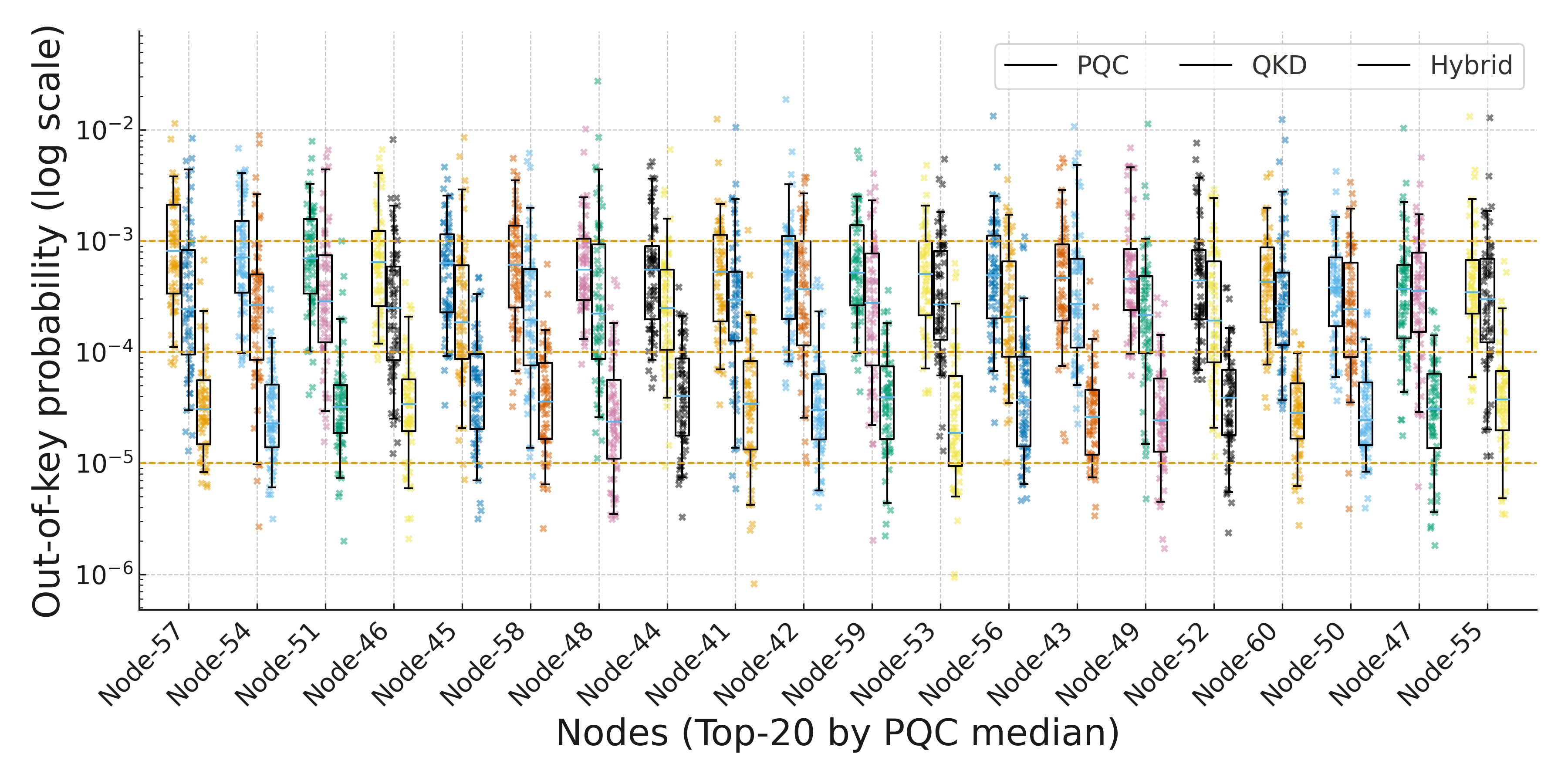}
  \caption{Top-20 stress nodes (selected by PQC median) out-of-key probability on a log scale; box summaries (5--95\% whiskers) with beeswarm of per-run samples.}
  \label{fig:pout_box_swarm}
\end{figure}

The node-centric view in Fig.~\ref{fig:pout_box_swarm} links service outcomes to key continuity. On the top-20 stress nodes, the median out-of-key probability drops from \(5.0\times 10^{-4}\) (\emph{PQC}) to \(2.6\times 10^{-4}\) (\emph{QKD}) and to \(3.1\times 10^{-5}\) (\emph{Hybrid}); at the 90th percentile the sequence is \(2.2\times 10^{-3}\), \(1.3\times 10^{-3}\), and \(1.4\times 10^{-4}\), respectively. This order-of-magnitude tail tightening under \emph{Hybrid} explains both the thinner availability lower tails and the reduced delay exceedance: a PQC fallback absorbs rare key-rate dips while preserving the lower medians associated with QKD-refreshed symmetric operation.

\subsection{Techno–economic Analysis}
\begin{figure}[t]
  \centering
  \includegraphics[width=\linewidth]{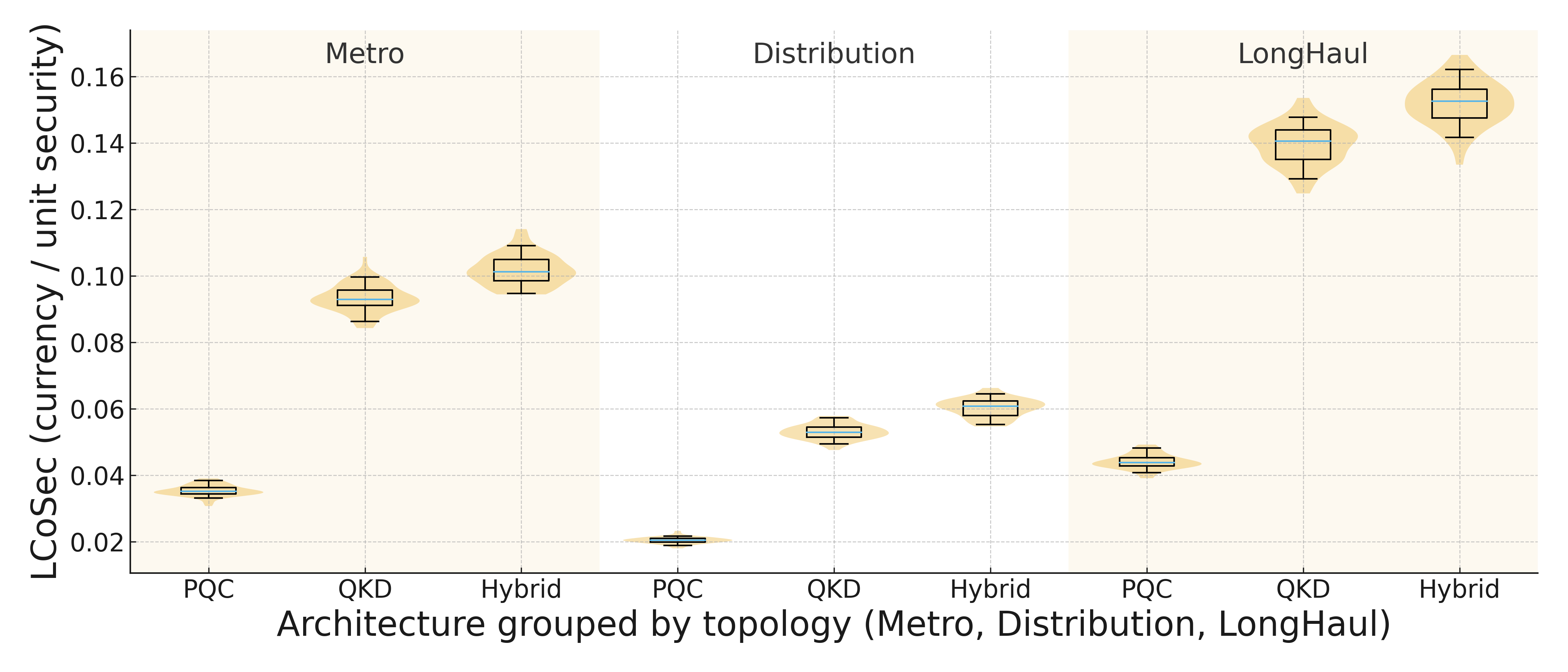}
  \caption{Levelized Cost of Security (LCoSec) distributions by architecture, grouped into Metro, Distribution, and LongHaul bands. Violin shapes depict full distributions and box overlays show 5--95\% whiskers.}
  \label{fig:te_lcosec}
\end{figure}

Figure~\ref{fig:te_lcosec} compares the levelized cost of security (LCoSec) across architectures and topologies. As expected, \emph{PQC-only} yields the lowest LCoSec in all topologies, reflecting its minimal capital intensity; however, LCoSec escalates markedly in the LongHaul case for both \emph{QKD-only} and \emph{Hybrid}, where trusted-node spans and higher lease/maintenance burdens dominate. The medians trace a consistent ordering: in Metro, \(\mathrm{LCoSec}_{\mathrm{PQC}}\approx 0.035\), \(\mathrm{LCoSec}_{\mathrm{QKD}}\approx 0.093\), and \(\mathrm{LCoSec}_{\mathrm{Hybrid}}\approx 0.102\); in Distribution, the triplet is roughly \(0.020/0.054/0.061\); in LongHaul, it rises to about \(0.044/0.151/0.156\) (currency per unit of effective security). The hybrid distributions are slightly broader than QKD in Metro and Distribution, consistent with the added but variable fallback overheads, while in LongHaul the two are statistically close due to the common long-distance optical cost drivers.

\begin{figure}[t]
  \centering
  \includegraphics[width=\linewidth]{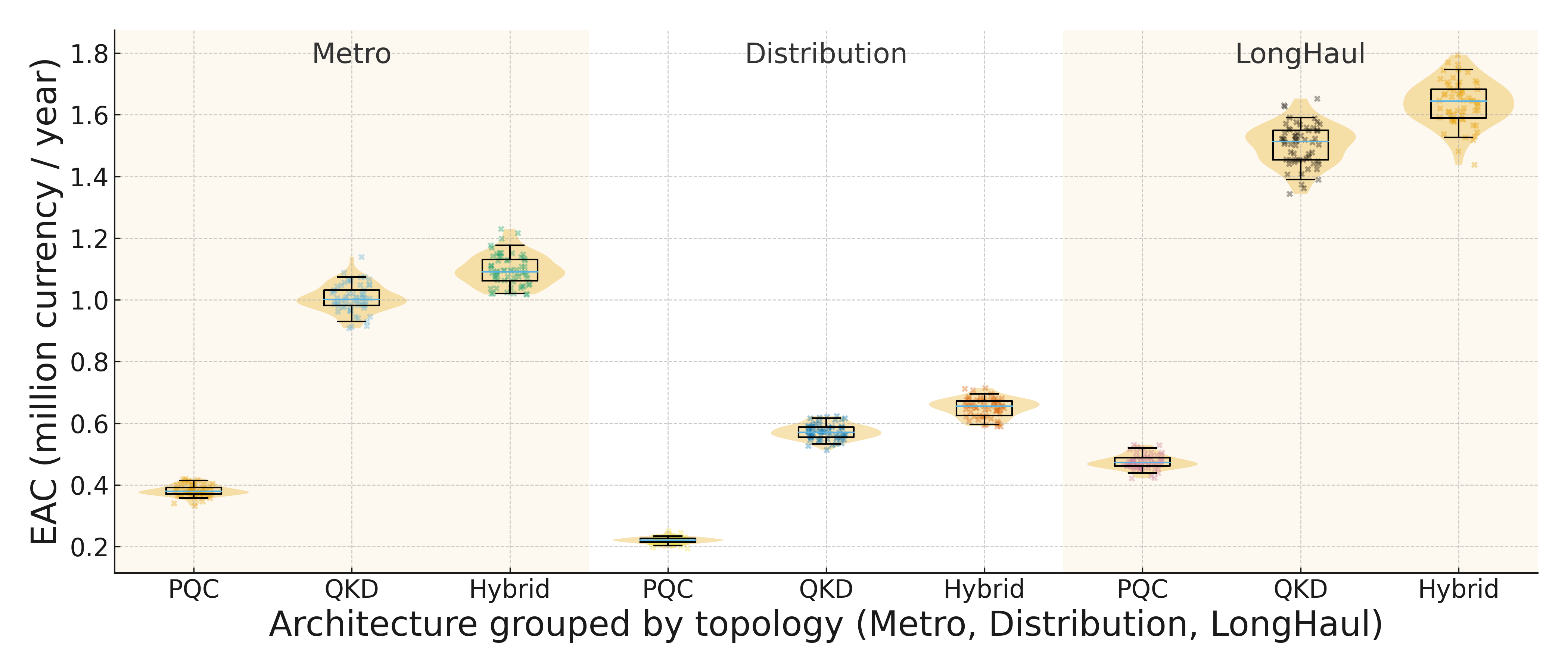}
  \caption{Equivalent annual cost (EAC) across architectures with a raincloud-style view (violin + box + jitter) and topology grouping.}
  \label{fig:te_eac}
\end{figure}

The raincloud view in Fig.~\ref{fig:te_eac} provides the annualized cost decomposition at a glance. Median EAC increases from \emph{PQC} to \emph{QKD} to \emph{Hybrid} in every topology, with the separation most visible in LongHaul where the interquartile band for \emph{Hybrid} sits around \(1.6\) million currency/year versus about \(1.5\) for \emph{QKD} and \(0.47\) for \emph{PQC}. The jittered points reveal moderate dispersion driven by OPEX and risk-cost variability, but without heavy tails—indicating that maintenance and SLA penalties, while material, are not the dominant uncertainty drivers compared to capital and lease rates.

\begin{figure}[t]
  \centering
  \includegraphics[width=\linewidth]{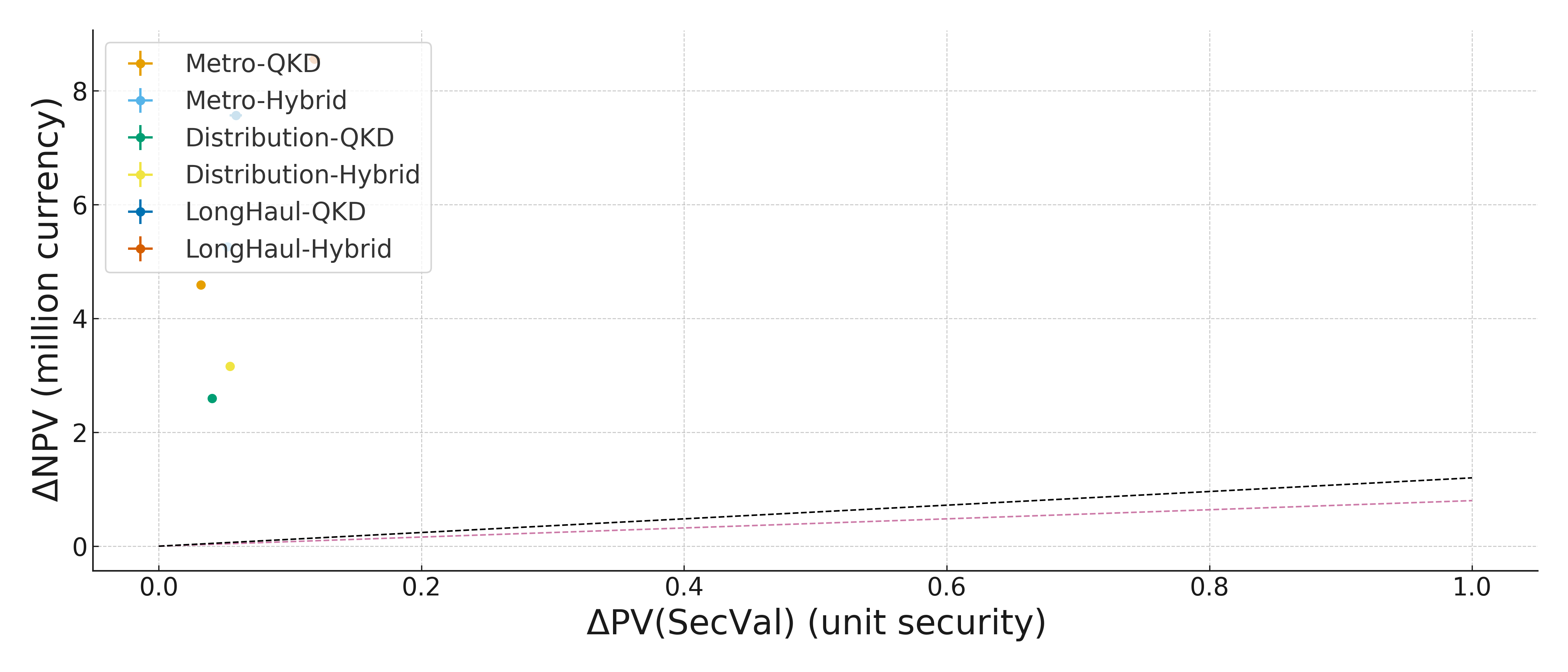}
  \caption{Incremental analysis relative to the PQC baseline: mean \(\Delta \mathrm{PV}(\mathrm{SecVal})\) versus \(\Delta \mathrm{NPV}\) with error bars (standard errors) for each topology and architecture. Dashed lines mark break-even loci \(\Delta \mathrm{NPV}=\pi\,\Delta \mathrm{PV}\) for \(\pi\in\{0.8,\,1.2\}\).}
  \label{fig:te_cis}
\end{figure}

Fig.~\ref{fig:te_cis} evaluates incremental efficiency relative to the PQC baseline. Across topologies, \emph{QKD} and \emph{Hybrid} both deliver positive but modest improvements in discounted effective security output \(\Delta \mathrm{PV}(\mathrm{SecVal})\), yet at substantially higher net present cost \(\Delta \mathrm{NPV}\). The Metro points lie closest to the \(\pi=0.8\) break-even line, implying that if the organization’s implicit value for a unit of effective security exceeds \(0.8\) (currency per unit), \emph{Hybrid} is near breakeven and \emph{QKD} remains marginal. Distribution sits between the two lines, suggesting that targeted use of \emph{Hybrid} could be justified by moderate shadow prices or regulatory incentives. LongHaul points lie well above both loci, indicating that under the assumed pricing and risk parameters, long-distance deployments would require either substantially higher valuation of security benefits, lower optical costs, or architectural refinements (e.g., fewer trusted regenerations) to cross the investment threshold. Overall, the three figures jointly indicate a clear techno–economic gradient: the relative appeal of \emph{Hybrid} improves as SLA risk and confidentiality valuation rise, but capital and long-haul optical penalties remain decisive in the levelized cost.

\subsection{Sensitivity and Robustness Analysis}
\begin{figure}[t]
  \centering
  \includegraphics[width=\linewidth]{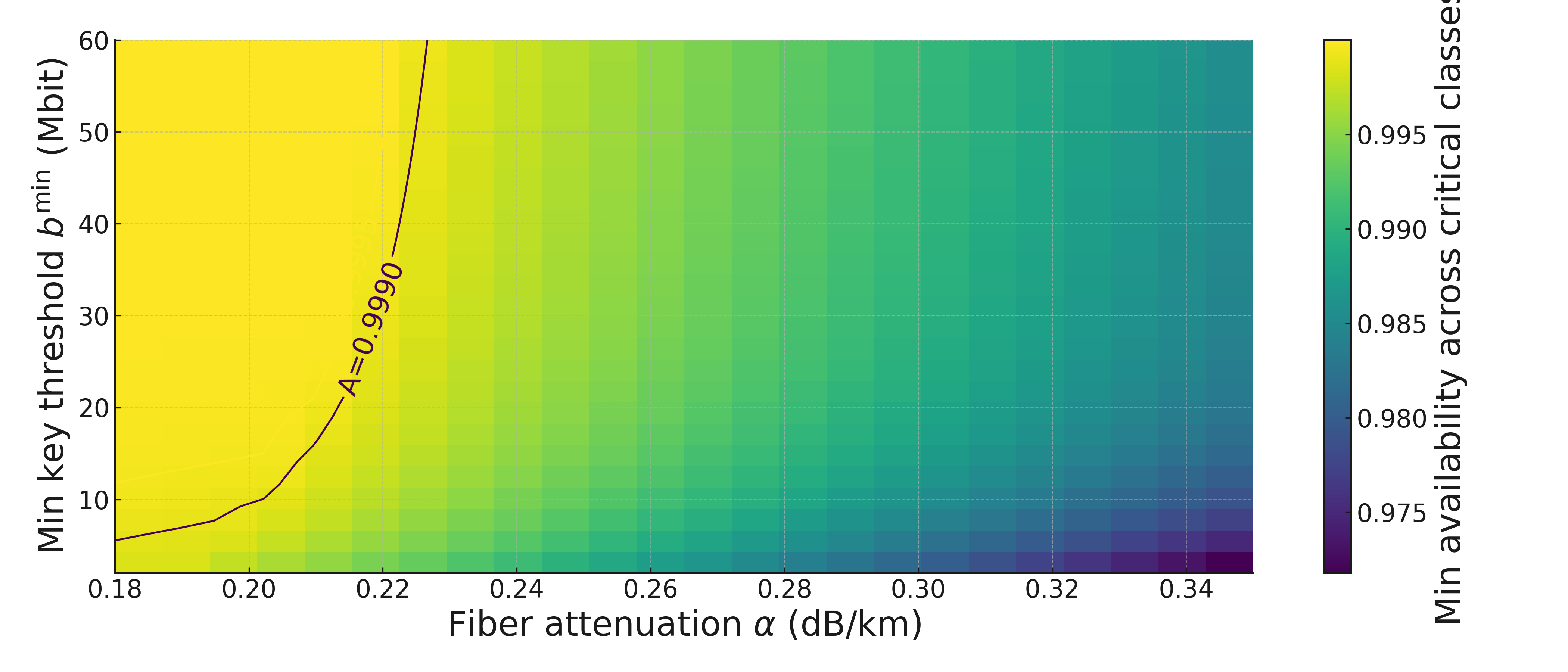}
  \caption{Hybrid architecture: heatmap of the minimum availability across critical classes as a function of fiber attenuation \(\alpha\) and minimum buffer threshold \(b^{\min}\); contours mark \(A=0.9990\) and \(A=0.9995\).}
  \label{fig:sr_heatmap_avail}
\end{figure}

Figure~\ref{fig:sr_heatmap_avail} reveals a clear trade-off between optical loss and buffer sizing. The \(A=0.9990\) contour tilts steeply with fiber attenuation \(\alpha\), indicating that every \(0.01\,\mathrm{dB/km}\) increase in \(\alpha\) must be compensated by a nontrivial rise in the safety threshold \(b^{\min}\) to keep the minimum availability above the SLA. In the low-loss region (\(\alpha\lesssim 0.21\,\mathrm{dB/km}\)), moderate buffers of \(10\text{--}15\,\mathrm{Mbit}\) suffice for \(A\ge 0.999\); once \(\alpha\) approaches \(0.28\,\mathrm{dB/km}\), the same target requires \(b^{\min}\gtrsim 35\,\mathrm{Mbit}\). The broad, warm plateau on the left shows diminishing returns: beyond roughly \(25\,\mathrm{Mbit}\), further increases in \(b^{\min}\) buy little improvement unless \(\alpha\) is simultaneously reduced, consistent with the exponential underflow sensitivity to buffer and the polynomial sensitivity to optical loss.

\begin{figure}[t]
  \centering
  \includegraphics[width=\linewidth]{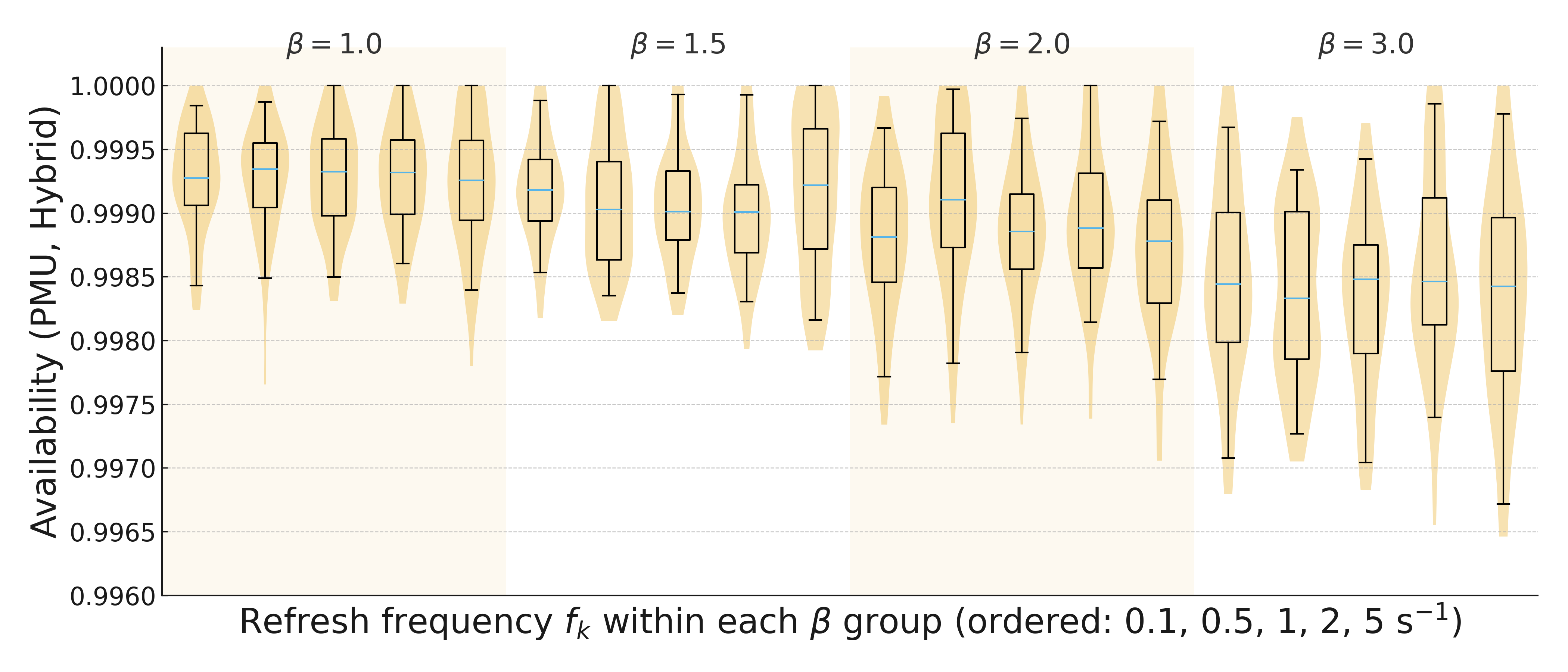}
  \caption{Hybrid (PMU): violin+box distributions of availability under refresh frequency \(f_k\in\{0.1,0.5,1,2,5\}\,\mathrm{s^{-1}}\) grouped by burst factor \(\beta_k\in\{1.0,1.5,2.0,3.0\}\); shaded bands separate \(\beta\) groups.}
  \label{fig:sr_violin_fb}
\end{figure}

Figure~\ref{fig:sr_violin_fb} studies parameter sensitivity at the crypto-policy layer. Raising the refresh frequency \(f_k\) from \(0.1\) to \(1\,\mathrm{s^{-1}}\) tightens and lifts the availability distributions, but very aggressive refresh (\(2\text{--}5\,\mathrm{s^{-1}}\)) begins to widen the lower tail, consistent with higher key draw and occasional fallback triggering under bursts. Increasing burst factor \(\beta_k\) systematically shifts the violins downward and broadens them, with the most pronounced degradation at \(\beta_k=3\). The pattern suggests a practical “sweet spot” at \(f_k\approx 1\,\mathrm{s^{-1}}\) for PMU traffic: it improves integrity without materially increasing key pressure across the tested burstiness range.

\begin{figure}[t]
  \centering
  \includegraphics[width=\linewidth]{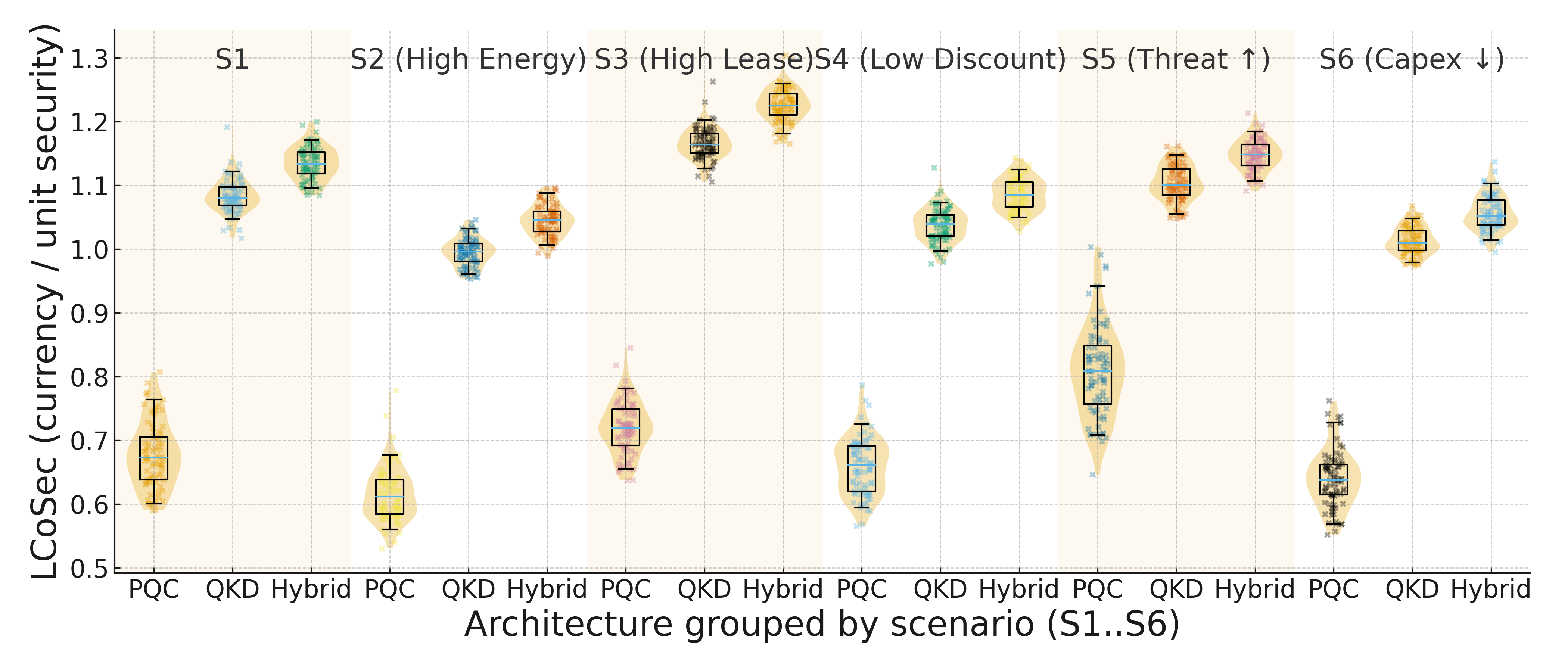}
  \caption{Scenario set \(\Omega\): raincloud-style view (violin+box+jitter) of LCoSec across architectures for six scenarios (baseline, high energy, high lease, low discount, threat escalation, capex reduction); shaded bands separate scenarios.}
  \label{fig:sr_scenario_raincloud}
\end{figure}

Robustness to exogenous uncertainty is summarized in Fig.~\ref{fig:sr_scenario_raincloud}. LCoSec distributions shift upward in scenarios with higher operating prices (S2) and higher lease burdens (S3), whereas a lower discount rate (S4) compresses the spread by reducing the relative weight of out-year OPEX and risk. Under threat escalation (S5), \emph{PQC} exhibits the largest uplift in LCoSec due to increased SNDL risk, while \emph{QKD} and \emph{Hybrid} move comparatively little. Conversely, a \(15\%\) capex reduction (S6) yields the largest gain for \emph{QKD} and \emph{Hybrid}, pulling their violins closer to the \emph{PQC} baseline. Across all scenarios, the jittered clouds remain compact without heavy tails, indicating that techno–economic variability is dominated by deterministic scenario shifts rather than high-variance residuals. Together, these results argue for a two-lever strategy: mitigate optical loss (route/plant choices) and maintain adequate buffers to secure SLA margins, while using moderate refresh frequencies to avoid avoidable key-pressure; on the economic side, long-life optical capex and lease rates are the principal robustness drivers, with threat inflation primarily penalizing \emph{PQC}-only designs.

\section{Conclusion}
This paper developed a unified techno-economic framework to evaluate the feasibility of quantum key distribution for secure power-system communications. The proposed stochastic model captured the joint dynamics of key demand, QKD supply, buffering, and fallback mechanisms, and provided analytical conditions for service-level assurance. Building on this model, two quantitative metrics, the Levelized Cost of Security (LCoSec) and the Cost of Incremental Security (CIS), were introduced to integrate capital, operational, and risk-related expenditures within a net-present-value formulation. Simulation studies across metropolitan, distribution, and long-haul topologies demonstrated that hybrid architectures combining QKD and post-quantum cryptography can significantly enhance availability and reduce tail risks in high-real-time and long-term confidentiality services. The results also revealed that the economic viability of QKD-enhanced systems depends strongly on fiber attenuation, buffer design, and key lifecycle policies. The findings provide quantitative evidence and practical guidelines for future deployment of QKD-based communication infrastructures in power systems, supporting risk-aware investment and policy decisions in the post-quantum era.

\bibliographystyle{IEEEtran}
\bibliography{refs}

\begin{thebibliography}{10}
\providecommand{\url}[1]{#1}
\csname url@samestyle\endcsname
\providecommand{\newblock}{\relax}
\providecommand{\bibinfo}[2]{#2}
\providecommand{\BIBentrySTDinterwordspacing}{\spaceskip=0pt\relax}
\providecommand{\BIBentryALTinterwordstretchfactor}{4}
\providecommand{\BIBentryALTinterwordspacing}{\spaceskip=\fontdimen2\font plus
\BIBentryALTinterwordstretchfactor\fontdimen3\font minus \fontdimen4\font\relax}
\providecommand{\BIBforeignlanguage}[2]{{%
\expandafter\ifx\csname l@#1\endcsname\relax
\typeout{** WARNING: IEEEtran.bst: No hyphenation pattern has been}%
\typeout{** loaded for the language `#1'. Using the pattern for}%
\typeout{** the default language instead.}%
\else
\language=\csname l@#1\endcsname
\fi
#2}}
\providecommand{\BIBdecl}{\relax}
\BIBdecl

\bibitem{Cao2022QKDNetworks}
\BIBentryALTinterwordspacing
Y.~Cao, Y.~Zhao, Q.~Wang, J.~Zhang, S.~X. Ng, and L.~Hanzo, ``The evolution of quantum key distribution networks: On the road to the qinternet,'' \emph{IEEE Communications Surveys \& Tutorials}, vol.~24, no.~2, pp. 839--894, 2022. [Online]. Available: \url{https://doi.org/10.1109/COMST.2022.3144219}
\BIBentrySTDinterwordspacing

\bibitem{Zhang2022DIQKD}
\BIBentryALTinterwordspacing
W.~Zhang, T.~van Leent, K.~Redeker, R.~Garthoff, R.~Schwonnek, F.~Fertig, S.~Eppelt, W.~Rosenfeld, V.~Scarani, C.~C.-W. Lim, and H.~Weinfurter, ``A device-independent quantum key distribution system for distant users,'' \emph{Nature}, vol. 607, pp. 687--691, 2022. [Online]. Available: \url{https://www.nature.com/articles/s41586-022-04891-y}
\BIBentrySTDinterwordspacing

\bibitem{Dolphin2023HybridChipQKD}
\BIBentryALTinterwordspacing
J.~A. Dolphin, T.~K. Para{\"i}so, H.~Du, R.~I. Woodward, D.~G. Marangon, A.~J. Shields \emph{et~al.}, ``A hybrid integrated quantum key distribution transceiver chip,'' \emph{npj Quantum Information}, vol.~9, 2023. [Online]. Available: \url{https://www.nature.com/articles/s41534-023-00751-3}
\BIBentrySTDinterwordspacing

\bibitem{Yang2024IntercitySPS}
\BIBentryALTinterwordspacing
J.~Yang, Z.~Jiang, F.~Benthin, J.~Hanel, T.~Fandrich, R.~Joos, S.~Bauer, S.~Kolatschek, A.~Hreibi, E.~P. Rugeramigabo, M.~Jetter, S.~L. Portalupi, M.~Zopf, P.~Michler, S.~K{\"u}ck, and F.~Ding, ``High-rate intercity quantum key distribution with a semiconductor single-photon source,'' \emph{Light: Science \& Applications}, vol.~13, 2024. [Online]. Available: \url{https://www.nature.com/articles/s41377-024-01488-0}
\BIBentrySTDinterwordspacing

\bibitem{Zahidy2024HiDQKD}
\BIBentryALTinterwordspacing
M.~Zahidy, D.~Ribezzo, C.~D. Lazzari, I.~Vagniluca, N.~Biagi, R.~M{\"u}ller, T.~Occhipinti, L.~K. Oxenl{\o}we, M.~Galili, T.~Hayashi, D.~Cassioli, A.~Mecozzi, C.~Antonelli, A.~Zavatta, and D.~Bacco, ``Practical high-dimensional quantum key distribution protocol over deployed multicore fiber,'' \emph{Nature Communications}, vol.~15, 2024. [Online]. Available: \url{https://www.nature.com/articles/s41467-024-45876-x}
\BIBentrySTDinterwordspacing

\bibitem{Liu2023PRL}
Y.~Liu \emph{et~al.}, ``{[Title not provided in user list]},'' \emph{Physical Review Letters}, vol. 130, p. 210801, 2023.

\bibitem{Wang2022NatPhotonics}
S.~Wang \emph{et~al.}, ``{[Title not provided in user list]},'' \emph{Nature Photonics}, vol.~16, pp. 154--161, 2022.

\bibitem{Chen2021npjQI134}
T.-Y. Chen \emph{et~al.}, ``{[Title not provided in user list]},'' \emph{npj Quantum Information}, vol.~7, p. 134, 2021.

\bibitem{Paraiso2021NatPhotonics}
T.~K. Para{\"i}so \emph{et~al.}, ``{[Title not provided in user list]},'' \emph{Nature Photonics}, vol.~15, pp. 850--856, 2021.

\bibitem{Erkilic2025CommunPhys}
O.~Erk{\i}l{\i}c \emph{et~al.}, ``{[Title not provided in user list]},'' \emph{Communications Physics}, vol.~8, p. 406, 2025.

\bibitem{Chen2021Nature}
Y.-A. Chen \emph{et~al.}, ``{[Title not provided in user list]},'' \emph{Nature}, vol. 589, pp. 214--219, 2021.

\bibitem{Xu2020RMP}
F.~Xu, X.~Ma, Q.~Zhang, H.-K. Lo, and J.-W. Pan, ``{[Title not provided in user list]},'' \emph{Reviews of Modern Physics}, vol.~92, p. 025002, 2020.

\bibitem{Pirandola2020AOP}
S.~Pirandola \emph{et~al.}, ``{[Title not provided in user list]},'' \emph{Advanced Optical Photonics}, vol.~12, pp. 1012--1236, 2020.

\bibitem{Alshowkan2022SciRep}
M.~Alshowkan \emph{et~al.}, ``{[Title not provided in user list]},'' \emph{Scientific Reports}, vol.~12, p. 12731, 2022.

\bibitem{Grice2025IEEEAccess}
W.~P. Grice \emph{et~al.}, ``{[Title not provided in user list]},'' \emph{IEEE Access}, vol.~13, pp. 17\,398--17\,413, 2025.

\bibitem{Aquina2025EPJQT}
N.~Aquina \emph{et~al.}, ``{[Title not provided in user list]},'' \emph{EPJ Quantum Technology}, vol.~12, p.~51, 2025.

\bibitem{Chaturvedi2025SciRep}
R.~Chaturvedi \emph{et~al.}, ``{[Title not provided in user list]},'' \emph{Scientific Reports}, vol.~15, p. 32831, 2025.

\bibitem{Wang2021npjQI67}
L.-J. Wang \emph{et~al.}, ``{[Title not provided in user list]},'' \emph{npj Quantum Information}, vol.~7, p.~67, 2021.

\bibitem{Garms2024AQT}
L.~Garms \emph{et~al.}, ``{[Title not provided in user list]},'' \emph{Advanced Quantum Technologies}, vol.~7, p. 202300304, 2024.

\bibitem{Cirigliano2024npjQI}
L.~Cirigliano \emph{et~al.}, ``{[Title not provided in user list]},'' \emph{npj Quantum Information}, vol.~10, p.~44, 2024.

\end{thebibliography}


\newpage
\vfill

\end{document}